\documentclass[journal]{IEEEtran}
\ifCLASSINFOpdf
\else
\fi
\usepackage{amsmath}
\usepackage[short,nocomma]{optidef}
\usepackage[hidelinks]{hyperref} 

\usepackage{graphicx} 

\usepackage[caption=false]{subfig} 

\usepackage[table]{xcolor}

\usepackage{mathtools} 

\usepackage{amssymb}

\usepackage{multirow, booktabs}
\usepackage{url}

\usepackage{etoolbox}
\makeatother
\makeatletter
\patchcmd{\@maketitle}
{\addvspace{0.5\baselineskip}\egroup}
{\addvspace{-1\baselineskip}\egroup}
{}
{}
\makeatother

\usepackage[switch,pagewise]{lineno}

\usepackage{eurosym}

\usepackage{nomencl}
\makenomenclature
\renewcommand\nomgroup[1]{%
	\item[\bfseries
	\ifstrequal{#1}{I}{Indices}{%
		\ifstrequal{#1}{S}{Sets}{%
			\ifstrequal{#1}{P}{Parameters}{%
				\ifstrequal{#1}{V}{Variables}{}}}}%
	]}

\graphicspath {{Results/}}

\begin{document}
	%
	\title{Three-Layer Joint Distributionally Robust Chance-Constrained Framework for Optimal Day-Ahead Scheduling of E-mobility Ecosystem}
	
	\author{Mahsa~Bagheri~Tookanlou,~\IEEEmembership{Member,~IEEE,} S.~Ali~Pourmousavi,~\IEEEmembership{Senior Member,~IEEE,} Mousa Marzband, ~\IEEEmembership{Senior Member,~IEEE} 
		
		\thanks{Mahsa~Bagheri~Tookanlou (\href{mailto:mahsa.tookanlou@northumbria.ac.uk}{mahsa.tookanlou@northumbria.ac.uk}) is with Faculty of Engineering and Environment, Department of Maths, Physics and Electrical Engineering, Northumbria University Newcastle, the United Kingdom, and S.~Ali~Pourmousavi (\href{mailto: a.pourm@adelaide.edu.au}{mailto: a.pourm@adelaide.edu.au}) is with the School of Electrical and Electronic Engineering, University of Adelaide, Australia, and Mousa~Marzband (\href{mailto:mousa.marzband}{mailto:mousa.marzband@northumbria.ac.uk}) is with Faculty of Engineering and Environment, Department of Maths, Physics and Electrical Engineering, Northumbria University Newcastle, the United Kingdomand also with the center of research excellence in renewable energy and power systems, King Abdulaziz University, Jeddah, Saudi Arabia. This work is funded by PGR scholarship at Northumbria University.}
		
	}

	
	%
	
	\markboth{Submitted to the IEEE Transactions ON POWER SYSTEMS FOR REVIEW, 2021}%
	{S.~Ali~Pourmousavi \MakeLowercase{\textit{et al.}}: An Optimal Day-Ahead Scheduling Framework for E-Mobility Ecosystem Operation with Drivers' Preferences}


	\maketitle
	
	\begin{abstract}
		A high number of electric vehicles (EVs) in the transportation sector necessitates an advanced scheduling framework for e-mobility ecosystem operation as a whole in order to overcome range anxiety and create a viable business model for charging stations (CSs). The framework must account for the stochastic nature of all stakeholders' operations, including EV drivers, CSs, and retailers and their mutual interactions. In this paper, a three-layer joint distributionally robust chance-constrained (DRCC) model is proposed to plan grid-to-vehicle (G2V) and vehicle-to-grid (V2G) operation in day-ahead for e-mobility ecosystems. The proposed stochastic model does not rely on a specific probability distribution for stochastic parameters. To solve the problem, an iterative process is proposed using joint DRCC formulation. 
		To achieve computational tractability, the exact reformulation is implemented for double-sided and single-sided chance constraints (CCs). Furthermore, the impact of temporal correlation of uncertain PV generation on CSs operation is considered. A simulation study is carried out for an ecosystem of three retailers, nine CSs, and 600 EVs based on real data from San Francisco, the USA. The simulation results show the necessity and applicability of such a scheduling method for the e-mobility ecosystem in an uncertain environment, e.g., by reducing the number of unique EVs that failed to reach their destination from 272 to 61. 
		
	\end{abstract}
	
	
	\mbox{}
	\nomenclature[I]{$t$}{Index for hours}
	\nomenclature[I]{$e$,$i$,$r$}{Index for EVs, CSs, and retailers, respectively}
	
	\nomenclature[I]{$m,n$}{Index of distribution network nodes}
	\nomenclature[S]{$B,E,R,S,T,F_1, F_2$}{Sets of Nodes, EVs, retailers, CSs, hours, first optional trip time, and second optional trip time, respectively}
	
	\nomenclature[P]{$a$}{End of day target SOC}
	\nomenclature[P]{$b,c,d,f$}{Cost of battery degradation parameters}
	\nomenclature[P]{$\zeta_{t,e}$}{Shortest driving route to reach the destination directly from origin of EV $e$ at time $t$ (km)}
	
	\nomenclature[P]{$\overline{r}_i^\mathrm{PV}/\underline{r}_{i}^\mathrm{PV}$}{Maximum/Minimum ramping rates of PV generation}
	
	\nomenclature[P]{$\overline{E}_e$}{Capacity of EV $e$’s battery (kWh)}
	
	\nomenclature[P]{$\eta_{e}^{+}/\eta_{e}^{-}$}{Efficiency of EV $e$’s battery in G2V/V2G mode (p.u.)}
	
	\nomenclature[P]{$\Delta t$}{Time interval (s)}
	
	\nomenclature[P]{$\mathcal{O}_{t,e,i}$}{Shortest driving distance between origin of EV $e$ and CS $i$ at time $t$ (km)}
	
	\nomenclature[P]{$\widehat{\mathcal{O}}_{t,e,i}$}{Shortest driving distance between origin of EV $e$ and the nearest CS at $t$ (km)}
	
	\nomenclature[P]{$it^{CS}/it^{re}$}{Number of iterations in CS/retailer layer}
	
	\nomenclature[P]{$\overline{it}^{CS}/\overline{it}^{re}$}{Maximum number of iterations in CS/retailer layer}
	
	\nomenclature[P]{$\Xi^{EV}/\Xi^{CS}/\Xi^{re}$}{Ambiguity set in EV/CS/retailer layer}
	
	\nomenclature[P]{$\overline {\alpha}/ \underline{\alpha}$}{Profit margin of the retailer}
	
	\nomenclature[P]{$\mathcal{D}_{t,e,i}$}{Shortest driving distance between CS $i$ and destination of EV $e$ at time $t$ (km)}

	\nomenclature[P]{$\overline{SOC}_{e}$/$\underline{SOC}_{e}$}{Maximum/Minimum SOC of EV $e$ (p.u.)}
	
	\nomenclature[P]{$\overline{SOC}_{i}^\mathrm{ESS}$/$\underline{SOC}_{i}^\mathrm{ESS}$}{Maximum/Minimum SOC of ESS at CS $i$ (p.u.)}
	
	\nomenclature[P]{$\Sigma_{t,i}^\mathrm{PV}$}{Covariance of PV generation in CS $i$ at time $t$}
	
	\nomenclature[P]{$\tau_{t}$}{Covariance of wholesale  electricity market prices at time $t$}

	\nomenclature[P]{$\Sigma_{e}$}{Covariance of initial SOC of EV $e$}
	
	\nomenclature[P]{$SOC_{e}^\mathrm{end}$}{SOC of EV $e$ at the end of the day (p.u.)}
	
	\nomenclature[P]{$\gamma_e$}{Power consumed by EV $e$ per km (kWh/km)}
	\nomenclature[P]{$\vartheta_{e}$}{EV $e$'s driver preference for minimum cost reduction during G2V operation (\$)}
	\nomenclature[P]{$\mathcal{G}_{e}$}{EV $e$'s driver preference for minimum revenue increase during V2G operation (\$)}
	\nomenclature[V]{$\widehat{\rho}^{+}_{t,i}$/$\widehat{\rho}^{-}_{t,i}$}{Electricity price offered by the closest CS to EVs at time $t$ in G2V/V2G mode (\$/kWh)}
	
	\nomenclature[P]{$\overline{E}_i^\mathrm{CH}$}{Capacity of chargers at CS $i$ (kW)}
	\nomenclature[P]{$\overline{N}_{i}^\mathrm{CH}$}{Maximum number of chargers in CS $i$}
	
	\nomenclature[P]{$\epsilon_{th}^\mathrm{EV}/\epsilon_{th}^\mathrm{CS}/\epsilon_{th}^\mathrm{re}$}{Theoretical risk parameter in each layer}
	
	\nomenclature[P]{$\overline{E}^\mathrm{GU}_{i}$/$\overline{E}^\mathrm{PV}_{i}$}{Capacity of CGU/PV system at CS $i$ (kW)}

	\nomenclature[P]{$\overline{E}^\mathrm{ESS}_{i}$}{Capacity of ESS at CS $i$ (kW)}
	
	
	

	\nomenclature[P]{$\widetilde{\rho}_t^\mathrm{WM}$}{Mean wholesale electricity market price at time $t$ (\$/kWh)}
	
	\nomenclature[P]{$\overline{E}_i$}{Capacity of CS $i$ (kW)}
	
	\nomenclature[P]{$\widehat{\mathcal{DO}}_{t,e}$}{Driving distance of EV $e$ to closest CS at time $t$ (km)}
	
	\nomenclature[P]{$\mathcal{K}_{e}$}{EV $e$'s driver preference for maximum extra distance to lower the cost compared to minimum route (km)}

	\nomenclature[P]{$HV$}{Heat value of fuel on the operation of gas turbine-generator ($kWh/m^3$)}
	
	\nomenclature[P]{$\rho_t^{gas}$}{Natural gas price at time $t$ ($\$/m^3$)}
	
	\nomenclature[P]{$\eta^\mathrm{GU}_{i}$/$\eta_i^\mathrm{CH}$}{Efficiency of CGU/chargers at CS $i$ (p.u.)}
	
	\nomenclature[V]{$\widetilde{Y}^\mathrm{PV}_{t,i}$ }{Mean local PV generation of CS $i$ at time $t$ (kW)}
	
	\nomenclature[P]{$\overline{\rho}^-$/$\underline{\rho}^-$}{Maximum/Minimum electricity prices offered by CSs for V2G service ($\$/kWh$)}
	
	\nomenclature[P]{$b_{m,n}$/$g_{m,n}$}{Susceptance/conductance of distribution line between node $m$ and $n$}

	\nomenclature[P]{$\underline{\Delta V}_m$/$\overline{\Delta V}_m$}{Lower/Upper limit of voltage deviation at node $m$}
	
	\nomenclature[P]{$\overline{\rho}^\mathrm{re}$/$\underline{\rho}^\mathrm{re}$}{Maximum/Minimum electricity prices offered by retailers to CSs ($\$/kWh$)}


	\nomenclature[V]{$X_{t,e,i}^{+}$/$X_{t,e,i}^{-}$ }{Charging/Discharging power of EV $e$ at CS $i$ at time $t$ (kW)}
	
	\nomenclature[V]{$\Delta\hat{V}_{m,t}$}{Voltage magnitude deviation obtained from the lossless power flow solution at node $m$ at time $t$}

	\nomenclature[V]{$\rho^{+}_{t,i}$/$\rho^{-}_{t,i}$}{Electricity price offered by CS $i$ at time $t$ for charging/discharging EVs (\$/kWh)}
	
	
	\nomenclature[V]{$SOC_{t,e}$}{SOC of EV $e$ at time $t$ (p.u.)}
	
	\nomenclature[V]{$\Gamma_{t,e,i}$/$\Pi_{t,e,i}$}{Binary variable for CS $i$ for charging/discharging EV $e$ at time $t$}
	
	\nomenclature[V]{$\beta_{t,i,r}$}{Binary variable for retailer $r$ by CS $i$ at time $t$}
	
	
	\nomenclature[V]{$Y^\mathrm{re}_{t,i,r}$/$Q^\mathrm{re}_{t,i,r}$}{Active/Reactive power purchased/provided from/by retailer $r$ by CS $i$ at time $t$ (kW/kVar)}
	
	\nomenclature[V]{$Y^\mathrm{GU}_{t,i}$}{Power produced by CGU system of CS $i$ at time $t$ (kW)}

	\nomenclature[V]{${\rho}^\mathrm{re}_{t,r}$}{Electricity price sold to CSs by retailer $r$ at time $t$ (\$/kWh)}

	\nomenclature[V]{$\rho^\mathrm{AG}_{t,i}$}{Electricity price sold to the aggregator by CS $i$ at time $t$ (\$/kWh)}

	\nomenclature[V]{$Y^+_{t,i}$/$Y^-_{t,i}$}{Charging/Discharging power of ESS of CS $i$ at time $t$ (kW)}
	
	\nomenclature[V]{$\psi_{t,i}$}{Binary variable for charging/discharging ESS at CS $i$}

	\nomenclature[V]{$P_{t,r}^{\mathrm{WM}}$/$Q_{t,r}^\mathrm{WM}$}{Active/Reactive power purchased/provided from/by the wholesale market by retailer $r$ at time $t$ (kW/kVar)}

	\nomenclature[P]{$\overline{P}_{m,n,t}$/$\underline{P}_{m,n,t}$}{Maximum/Minimum active power flow between node $m$ and $n$ (kW)}
	
	\nomenclature[P]{$\overline{Q}_{m,n,t}$/$\underline{Q}_{m,n,t}$}{Maximum/Minimum reactive power flow between node $m$ and $n$ (kVar)}
	
	\nomenclature[V]{$P_{m,n,t}$/$Q_{m,n,t}$}{Active/Reactive power flow between node $m$ and $n$ at time $t$ (kW/kVar)}
	
	\nomenclature[V]{$\Delta V_{m,t}$}{Voltage magnitude deviation at node $m$ at time $t$}
	
	\nomenclature[V]{$\Delta \theta_{m,t}$}{Voltage angle deviation at node $m$ at time $t$}
	
	\nomenclature[V]{$V_{m,t}$}{Voltage magnitude at node $m$ and time $t$}
	
	\nomenclature[V]{$\theta_{m,t}$}{Voltage angle at node $m$ and time $t$}
	
	
	\nomenclature[P]{$VCS$}{Virtual charging station}
	
	\nomenclature[P]{$\mathbb P$}{Probability distribution}
	
	\nomenclature[V]{$A_{t,e,i}$}{Sum of charging and discharging power of EV $e$ in CS $i$ at time $t$}
	
	\nomenclature[V]{$\widehat A_{t,e,i}$}{Sum of charging and discharging power of EV $e$ in the nearest CS at time $t$}
	
	\nomenclature[P]{$\widehat X^{+}_{t,e,i}/\widehat X^{-}_{t,e,i}$}{Charging/discharging power of EV $e$ in the nearest CS at time $t$}

	\nomenclature[P]{$\widetilde{SOC}_{0_e}$}{Mean of initial SOC of EV $e$}
	
	\nomenclature[V]{$\omega_e$}{A random variable with zero mean and covariance matrix $\Sigma_e$ for EV $e$}
	
	\nomenclature[V]{$\kappa_{t,r}$}{A random variable with
		zero mean and covariance matrix $\tau_{t}$ for retailer $r$ at time $t$}

	\nomenclature[V]{$\xi_{t,i}$}{A random variable with
		zero mean and covariance matrix $\Sigma_{t,i}^{PV}$ for CS $i$ at time $t$}

	\printnomenclature

	\IEEEpeerreviewmaketitle

	\section{Introduction}
	\label{sec:introduction}
	
	\IEEEPARstart WITH the increasing adoption of electric vehicles (EVs) in the transportation sector and the rising number of charging stations (CSs) equipped with renewable generation resources, application of a coordinated vehicle-to-grid (V2G) and grid-to-vehicle (G2V) operation of e-mobility ecosystems under inherent uncertainties has become inevitable. While day-ahead scheduling can reduce range anxiety of the drivers, using a deterministic model to do so might not fulfil the actual driving requirements of the EV drivers due to inaccurate estimation of stochastic parameters for the next day; hence leading to ineffective outcomes and drivers' disappointment. In fact, ignoring the impact of underlying uncertainties in an e-mobility ecosystem might result in significant losses for all stakeholders and might introduce concerns and challenges for power system operation. This indicates the importance of a scheduling framework that accounts for the different source of uncertainty in the e-mobility ecosystem operation. 
	
	The major sources of uncertainty in the future e-mobility ecosystem originate from the EV drivers behavior, unpredictable nature of renewable energy resources (RES) at the CSs and the wholesale electricity market prices \cite{zhao2018robust, hajebrahimi2020scenario}. 
	To consider these sources of uncertainties in the scheduling problem, 
	various approaches have been proposed in the literature including robust optimization and scenario-based stochastic programming, which are commonly used to characterise the uncertainties in transportation sector \cite{sun2020robust, zhou2017optimal, fallah2020charge}. However, each of these approaches poses certain challenges. For scenario-based methods, an adequate number of scenarios must be considered to sufficiently represent the stochasticity of the parameters. This is because the performance of stochastic models highly depends on the assumed scenarios. More than often, it requires extra computational time and sometimes results in computational intractability. In robust optimisation approaches, the worst-case scenario is considered, which may lead to the most conservative solutions \cite{roos2020reducing}. Also, it is challenging to define a proper probability distribution function for stochastic parameters. To properly address these challenges in stochastic programming, a distributionally robust chance-constrained (DRCC) model is proposed to consider an ambiguity set, which encompasses a family of probability distributions with the first- and second-order moments. Also, the ambiguity set with mean and covariance matrix obtained from empirical data can perfectly describe the temporal correlation of uncertainties. In the following subsections, a comprehensive literature review is presented followed by a statement listing the contributions of this paper.
	\par
	
	\subsection{Literature review}
	\label{subsec:literature review}
	Numerous research papers in the literature focused on EVs' G2V and V2G scheduling in an uncertain environment in recent years. We categorize the existing studies based on different methods used, namely (1) scenario-based methods, (2) robust optimization-based methods, and (3) chance-constrained (CC) optimization-based approaches. The first group of studies focused on scenario-based methods using stochastic programming for e-mobility ecosystem operation. For instance, a scenario-based stochastic approach was proposed 
	in \cite{tian2020risk} to obtain optimal scheduling of plug-in EVs for aggregators by maximizing their profit in day-ahead and reserve markets. To consider uncertainties, the risk-constrained stochastic optimization model was proposed in that paper. In \cite{wang2020scenario}, a scenario-based two-stage stochastic problem was offered based on a rolling window approach for scheduling EVs in G2V operation for different grid requirements. The goal was to minimize the distance between the actual and target state of charge (SOC) over a given time span. The arrival and departure times, and the initial and target SOC of EV batteries were considered as uncertain parameters.
	A two-stage stochastic model was developed in \cite{schucking2021two} to optimize the investment decision and the operational cost of EVs in the first and second stage, respectively, considering energy consumption and available charging times as the sources of uncertainties. The stochastic study was investigated based on different scenarios, which was generated by a hidden Markov model. In \cite{zhang2013charging}, an EV charging scheduling model was presented to minimize the mean waiting time of EVs at a CS with multiple charging points equipped with RES. The EV arrival, the intermittency of the RES, and the electricity price were considered as uncertain parameters and described as independent Markov processes. In \cite{wu2020real}, optimal control of a CS with a PV system was investigated based on a finite-horizon Markov decision model under uncertainties of EV drivers' behaviors and dynamic electricity prices. Then, the total operation cost of the CS was minimized considering the V2G service and battery degradation. In \cite{faridimehr2018stochastic}, a two-stage scenario-based stochastic framework was developed for modeling the optimal network of CSs aiming to find the optimal CS for plug-in hybrid EVs. In that study, stochastic parameters were the battery demand, initial SOC, preferences for charging, and RES generation. In the first stage, the deterministic problem was solved leading into the second stage, where the decision was made considering the uncertainties. In \cite{liu2017dynamic}, a dynamic stochastic optimization problem was solved to determine optimal EV charging cost considering electricity prices, RES production, and load as stochastic parameters. The authors in \cite{wang2016predictive} proposed a predictive framework by accounting for the uncertainties of EV drivers' behaviors to achieve cost-efficient solutions. A kernel-based method was used to estimate uncertain parameters in G2V operation.  \par

	The second group of studies explored the application of robust optimization in the e-mobility ecosystem. For example, in \cite{zeng2020bi}, a bi-level robust optimization model was formulated to optimize the design of a CS considering uncertainties in real-time operation of the CS including the electricity prices, RES, and the number of EVs. In \cite{sun2020robust}, a robust day-ahead scheduling approach was developed for EV charging in a stochastic environment in order to simultaneously deal with EV drivers' requirements and distribution network constraints. Several uncertainties were considered including daily trip distances and arrival and departure times. Furthermore, conservative day-ahead assumptions were considered in the proposed model to address the negative effects of uncertainties. In \cite{bai2015robust}, a robust optimization-based unit commitment model was developed for a system with thermal generators and EV aggregators in day-ahead V2G planning problem. The robust optimization model was then reformulated as a deterministic mixed-integer quadratic program using explicit maximization method. The uncertain parameter was the available energy capacity of each EV aggregator. In \cite{yang2015noncooperative}, a robust Stackelburg game was used to investigate the interaction between an aggregator as the leader and several plug-in hybrid EVs as the followers for charging scheduling under energy demand uncertainty. Cooperative and non-cooperative games were investigated to analyze charging scheduling in that study, and selling electricity price to EVs was determined by maximizing the utility of the aggregator. Then, the EV schedules were obtained by optimizing the utility function of EVs.
	A deterministic optimization problem for optimal charging schedules and its robust counterparts under uncertainty of electricity prices were compared in \cite{korolko2015robust}. Trade-offs between optimality of the cost function and robustness of charging scheduling were investigated and stability of robust charging schedules were obtained with respect to uncertain electricity prices. In \cite{cao2020optimal}, robust scheduling of EV aggregators operation was investigated under uncertainty of electricity prices to maximize its profit. In \cite{cui2020optimal}, a Stackelberg game was proposed for the EV aggregator (as the leader) and EVs (as followers) to determine optimal day-ahead charging and frequency reserve scheduling aiming to balance the benefits of the players in the game. The robust optimization approach was used to investigate EVs' optimal schedules under uncertain frequency regulation signals. \par

	The application of CC optimization has been investigated in several studies in this field. For instance, to consider the stochastic features of the EV driving patterns in \cite{liu2016distribution}, a CC optimization problem of EV aggegators and the distribution system operator were developed as a mixed-integer quadratic program. The goal was to reduce the congestion in the distribution network with a large amount of EVs. In \cite{liu2019day}, the CC programming was used to develop a day-ahead scheduling strategy for an EV battery swapping station. EV's battery swapping and PV generation were considered as the source of uncertainties, described by a probabilistic sequence. In \cite{wang2019chance}, a two-stage program for energy management system of the distribution networks with EVs and RES was presented. In the first stage, a CC model was developed to obtain the optimal operation of CSs and battery swapping stations under uncertainties of RES generation. In the second stage, the regulated EV charging was determined to meet the EV's charging demand following the optimal operation of CSs.

	\par
	Based on the comprehensive literature review presented above, we identified four gaps in knowledge concerning EV scheduling in an e-mobility ecosystem 
	as follows:
	
	\begin{itemize}
		
		\item The interactions between stochastic parameters originated from different stakeholders in an e-mobility ecosystem was ignored by optimizing each stakeholder’s operation individually; 
		
		\item Specific probability distribution functions were assumed for modelling of stochastic parameters, which mostly are just an estimation of true underlying model \cite{pourahmadi2019distributionally};
		
		\item The proposed CC models treated the lower and upper bounds as two single-sided CCs, which may lead to over- or under-estimation of the parameters; hence violating the constraints in reality; 
		
		\item The impacts of temporal correlation of PV generation uncertainty on CSs operation has not been investigated in a chance-constraint formulation in this field.

	\end{itemize}
	
	\subsection{Main Contributions}
	\label{subsec:contributions}
	
	In this paper, a three-layer joint DRCC framework is proposed to schedule V2G and G2V operation in the day ahead for an e-mobility ecosystem including EVs, CSs, and retailers in an uncertain environment with unknown probability distributions. In an attempt to facilitate the investigation of an uncertain e-mobility ecosystem, the interactions between stochastic nature of the three stakeholders are considered in the proposed model. A family of probability distributions with the same mean and covariance matrix is defined, called a moment-based ambiguity set, to solve a stochastic program without relying on a specific distribution function. An exact second-order cone programming reformulation of joint DRCC day-ahead scheduling framework is developed, which ensures that violation of both upper and lower limits of a constraint remains small for the worst-case probability under the ambiguity set. Furthermore, the temporal correlation of the PV system generation in each time interval is considered in the joint DRCC model. \par
	
	
	
	


	\par
	The rest of this paper is structured as follows: Section~\ref{sec:problem} presents problem definition and describes the stochastic G2V and V2G framework including the three stakeholders. It is followed by the proposed three-layer joint DRCC formulation in Section~\ref{sec: modeling framework}. In Section~\ref{sec: simulation results}, an ecosystem based on 600 EVs, nine CSs, and three retailers is devised for simulation study and the results are discussed. The conclusion remarks are presented in Section~\ref{sec: conclusion}. An electronic companion is produced in \cite{electronic} to explain and demonstrate reformulation of the single-sided and double-sided CCs in EV, CS, and Retailer layers. 
	
	\section{Problem Definition}
	\label{sec:problem}
	This paper offers a three-layer joint DRCC scheduling framework in the day ahead, 
	where the DRCC model of each stakeholder is developed in each layer. 
	We model the uncertainty of each player by means of an ambiguity set, which consists of a family of probability distributions of each uncertain parameter with the first- and second-order moments, i.e, mean and covariance of available historical data. In the proposed ecosystem, we consider $R$ number of retailers, indexed by $r \in \{1,2,...,R\}$. Retailers sell electricity to CSs from the wholesale electricity market. Therefore, wholesale electricity price at time $t$ is the major source of uncertainty in this layer. There are $S$ number of CSs in the ecosystem, indexed by $i \in \{1, 2, ..., S\}$, physically located in the scheduling area. They operate at the distribution network level and provide V2G and G2V services to EVs. Without loss of generality, it is assumed that each CS possesses small gas turbine/diesel generator as a conventional generation unit (CGU), PV, and energy storage system (ESS) to supply electricity to EVs during G2V operation. Also, CSs purchase electricity from EVs and sell it in the wholesale electricity market through aggregators \cite{tookanlou2021optimal}. PV generation is the main source of uncertainty in the CS layer. During a typical day, it is assumed that EVs can have multiple mandatory trips and optional trips, which are explained in detail in our previous paper \cite{tookanlou2021optimal}. 
	In this study, EVs have two mandatory trips, which must be fulfilled at any cost. However, as opposed to mandatory trips, an optional trip is selected if it offers a possibility for EV drivers to benefit from cheap G2V or expensive V2G services outside of the mandatory trip hours. Each EV driver may select only one or two optional trips during a day to reduce its cost. EVs with a known location and initial SOC, which is the source of uncertainty in EV layer, seek G2V and V2G plans for the combined mandatory and optional trips such that it minimizes their overall cost while fulfilling their preferences. A day before the scheduling day, EV drivers send their trip plans to the scheduling centre. Then, the scheduling problem is solved for the entire day ahead. The driving routes between CS $i$ and origin of EV $e$ in each trip are known and only one of CSs might be selected for EV $e$. Thus, two binary variables are assigned to each CS for G2V and V2G operation of the EV $e$ at each time interval. The virtual CS (VCS) is considered for the case in which the most economic decision for EV $e$ is not to be charged nor discharged in a trip. Thus by selecting a VCS, EV $e$ reaches to the destination from its origin without charging or discharging, while the EV’s preferences and constraints are satisfied. 
	Therefore, in a mandatory trip, charging and discharging power of a VCS are equal to zero for EV $e$. The only difference between mandatory and optional trips is that, the driving route of a VCS is considered zero for the optional trips. Please see \cite{tookanlou2021optimal} for further details on this modelling approach.

	\section{Modeling Framework}
	\label{sec: modeling framework}
	
	In this section, the day-ahead joint DRCC V2G and G2V scheduling framework, which includes several uncertain and deterministic constraints, is presented for each layer based on the deterministic model developed by the authors in \cite{tookanlou2021optimal}. A general formulation looks like Eq.~\ref{eq: general CC}, which finds minimizers $x$ to $f:\mathbb{R}^n \longrightarrow \mathbb{R}$ as the objective function of a CC program subject to a set of deterministic constraints ($D$) and stochastic constraints ($H\left(x,\lambda\right)\leq 0, H: \mathbb{R}^n \times \Lambda \longrightarrow\mathbb{R}^m$). $H$ is needed to satisfy any probability distribution ($\mathbb{P}$) from the ambiguity set ($\Xi$) at a given confidence level $(1-\epsilon) \in (0,1)$.
	\begin{equation}
		c^*=min \bigg\{f(x): x\in D,
		\inf_{\mathbb{P}\in \Xi 
		} \mathbb{P} \big[\lambda \in \Lambda : \{ H(x,\lambda)\leq 0\}\big]\geq1-\epsilon
		\bigg\}.
		\label{eq: general CC}
	\end{equation}
	
	In the joint DRCC, the uncertain parameter is modeled by $\mu+\omega$, where $\mu$ is the first-order mean of the uncertain parameter and $\omega$ is a random variable with zero mean and covariance matrix $\Sigma$. The ambiguity set is defined by
	
	\begin{equation}
		\Xi=\big\{\mathbb{P} \in \Xi(\mathbb{R}^\upsilon): \mathbb{E}_{\mathbb{P}}[\omega], E_{\mathbb{P}}[\omega \omega^T]=\Sigma\big\}.
		\label{eq: ambiguity set}
	\end{equation}
	
	The DRCC models for EV, CS, and Retailer layers are presented in Section~\ref{subsec:DRCC in EV}, ~\ref{subsec:DRCC in CS}, and ~\ref{subsec:DRCC in Retailer}, respectively. The exact reformulation of the single-sided and double-sided CCs are presented in Theorem 1 and Theorem 2 in \cite{xie2017distributionally}. The reformulation of our problem are described in the electronic companion in \cite{electronic}. \par
	
\begin{figure}[!htb]
		\setlength\abovecaptionskip{-0.5\baselineskip}
		\centering
		\includegraphics[ width=1\columnwidth]{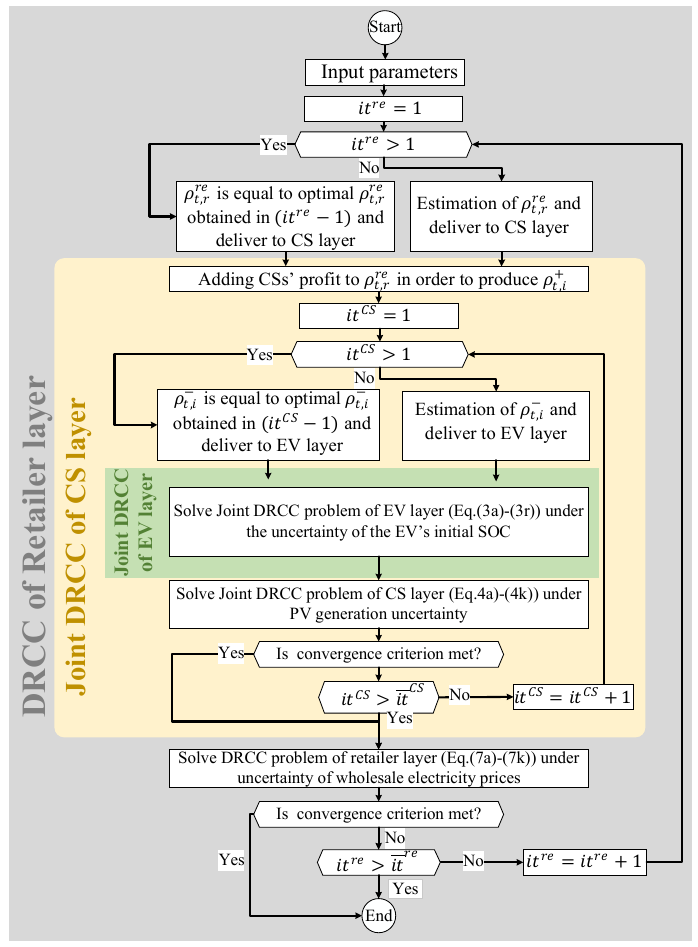}
		\caption{Flowchart of the three-layer joint DRCC problem}
		\label{fig: flowchart}
		\captionsetup{justification=centering}
	\end{figure}

	The proposed three-layer scheduling framework is solved by an iterative approach shown in Fig.~\ref{fig: flowchart}. The retailers estimate wholesale electricity prices using historical data in the first iteration. The prices are passed on to the CS layer. In this iteration, the prices are only inflated to consider the CSs’ profit margin. Then, the CSs’ will determine their prices accordingly and pass them on to EV layer, where the first DRCC problem will be solved in the first iteration. Also, the V2G prices are estimated by the CS layer in the first iteration. In the EV layer, under the uncertainty of the EV's initial SOC, the decision variables including EVs’ charging and discharging power and CS selection for each trip will be determined by solving the joint DRCC problem. The inner loop, which is between CS and EV layers, will continue until the convergence criterion of the DRCC problem in the CS layer under PV generation uncertainty is satisfied. Afterwards, selected retailers and the amount of purchased power from each retailer will be communicated back to the retailer layer. Then, the DRCC problem in the retailer layer is solved considering the uncertainty of the wholesale electricity prices. In retailer layer, new electricity prices offered by retailers to CSs will be determined according to the reactions from the CSs and EVs to the original prices. Second iteration of the outer loop starts with the new retailers’ prices. The iterative process will be terminated once the difference between the relevant objective functions in the last two iterations for both inner and outer loops is less than or equal to 0.001.
	
	
	\subsection{Joint DRCC Model in EV Layer}
	\label{subsec:DRCC in EV}
	The proposed DRCC model in EV layer includes Eqs.~\eqref{eq: Net cost of EVs}-\eqref{eq: Distance to VCS for the second optional trip} to minimise the net cost of EV operation during V2G and G2V services. The proposed problem constrained by a set of distributionally robust chance constraints in Eqs.~\eqref{eq: CC of SOC of EVs},~\eqref{eq: CC of SOC at the end of a day},~\eqref{eq: CC of Preference cost}, and ~\eqref{eq: CC of Preference revenue}. 
	Equations~\eqref{eq: sum of charging and discharging} and \eqref{eq: sum of charging and discharging at nearest EV} indicate sum of the charging and discharging power of EV $e$ in CS $i$ and the nearest CS, respectively, at time $t$. Equation~\eqref{eq: CC of SOC of EVs} signifies that under the worst distribution in ambiguity set for EV layer ($\Xi^{EV}$), the probability of maintaining the SOC level of EV $e$ within a lower and upper bound at all times must be greater than or equal to a given confidence level. The DRCC in Eq.~\eqref{eq: CC of SOC at the end of a day} points to fulfil target SOC of EV $e$ at the end of the day within a permissible range (i.e., between the target and maximum SOC). Admissible charging and discharging capacity of the chargers at CS $i$ are imposed by Eqs.~\eqref{eq: Limitation of charging power} and~\eqref{eq: Limitation of discharging power}. In order to select one CS for either G2V and V2G services by EV $e$ at time $t$, sum of the binary variables of CSs must be less than or equal to one, as in by Eq.~\eqref{eq: one mode of charging/discharging}. Furthermore, 
	Eq.~\eqref{eq: Limitation of chargers for charging and discharging} guarantees that the number of used chargers at a CS for charging and discharging does not exceed the number of existing chargers. The DRCC in Eqs.~\eqref{eq: CC of Preference cost} and ~\eqref{eq: CC of Preference revenue} impose driver's cost/revenue threshold limitations for selecting alternative route instead of the shortest route during V2G and G2V operation. The driver's route preferences for charging/discharging EV $e$ at time $t$ in an alternative CS other than the nearest CS are ensured by Eqs.~\eqref{eq: Distance preference for charging} and ~\eqref{eq: Distance preference for discharging}. The zero power of charging and discharging at a VCS must be set to zero, which is achieved by Eqs.~\eqref{eq: charging power of VCS} and ~\eqref{eq: discharging power of VCS}. Equation~\eqref{eq: Distance to VCS for mandatory trip} sets the driving route allocated to VCS for the mandatory trip, which is equal to the shortest route to reach the destination directly from origin of EV $e$. The driving route distance between EV and VCS in the first and second optional trips are set to zero by Eqs.~\eqref{eq: Distance to VCS for the first optional trip} and~\eqref{eq: Distance to VCS for the second optional trip}.

	\begin{mini!}|s|[3]<b>
		{\substack{X_{t,e,i}^{+},X^{-}_{t,e,i}, 
				\\ 
				\Gamma_{t,e,i}
				,\Pi_{t,e,i}
		}}{\sup_{\mathbb P\in \Xi^{EV}}\mathbb E_\mathbb P \bigg[\begin{aligned}[t]&{\sum\limits_{t=1}^{T}\sum\limits_{e=1}^{E}} X_{t,e,i}^{+}\cdot \rho_{t,i}^{+}\\ 
				&+\bigg(b\cdot{\big(SOC_{t,e}-a\cdot(\Gamma_{t,e,i}+\Pi_{t,e,i})\big)}^2\\
				&+c.X^{+}_{t,e,i}-d.X^{-}_{t,e,i}+{f}.{X^{-}}^2_{t,e,i}\bigg)\\
				&-X_{t,e,i}^{-}.\rho_{t,i}^{-}\bigg]\quad\quad\quad\forall i \in S,
		\end{aligned}}{\label{eq: Net cost of EVs}}{}
		\addConstraint{\begin{aligned}[t]
				A_{t,e,i}=\frac{\eta_{e}^{+}.\sum\limits_{t=1}^{t}X^{+}_{t,e,i}.\Delta t}{\overline{E}_{e}}-\frac{\sum\limits_{t=1}^{t}X^{-}_{t,e,i}\Delta t}{\eta_{e}^{-}\overline{E}_{e}},{\label{eq: sum of charging and discharging}}
		\end{aligned}}
		\addConstraint{\begin{aligned}[t]
				\widehat{A}_{t,e,i}=\frac{\sum\limits_{t=1}^{t}\eta_{e}^{+}.\widehat{X}^{+}_{t,e,i}\Delta t}{\overline{E}_{e}}-\frac{\sum\limits_{t=1}^{t}\widehat{X}^{-}_{t,e,i}\Delta t}{\overline{E}_{e}.\eta_{e}^{-}},{\label{eq: sum of charging and discharging at nearest EV}}
		\end{aligned}}
		\addConstraint{\begin{aligned}[t]{\inf_{\mathbb P\in \Xi^{EV}}\mathbb P}&\bigg[\underline{SOC}_{e}\leq\widetilde{SOC}_{0_e}-e^{T}{\omega_e}+A_{t,e,i}\\&{-\sum\limits_{t=1}^{t}{\frac{\zeta_{t,e}.\gamma_e}{\overline{E}_{e}}.(1-\Gamma_{t,e,i}-\Pi_{t,e,i})}}-\frac{\mathcal{O}_{t,e,i}.\gamma_e}{\overline{E}_{e}}.\\&(\Gamma_{t,e,i}+\Pi_{t,e,i})
				\leq \overline{SOC}_{e}\bigg]\geq1-\epsilon_{th}^{EV}\\&{\quad \forall t \in T, \forall e \in E, \forall i \in S},\end{aligned}}{\label{eq: CC of SOC of EVs}}
		\addConstraint{\begin{aligned}[t]{\inf_{\mathbb P\in \Xi^{EV}}\mathbb P}&\bigg[SOC_{e}^\mathrm{end} \leq\widetilde{SOC}_{0_e}-e^{T}\omega_e+A_{T,e,i}\\&{-\sum\limits_{t=1}^{T}{\frac{\zeta_{t,e}.\gamma_e}{\overline{E}_{e}}(1-\Gamma_{t,e,i}-\Pi_{t,e,i})}
					- \frac{\mathcal{O}_{T,e,i}.\gamma_e}{\overline{E}_{e}}}.\\&(\Gamma_{T,e,i}+\Pi_{T,e,i})
				\leq \overline{SOC}_{e}\bigg]\geq1-\epsilon_{th}^{EV} \\&{\quad \forall t \in T, \forall e \in E, \forall i \in S},\end{aligned}}{\label{eq: CC of SOC at the end of a day}}
		\addConstraint{0 \leq X^{+}_{t,e,i}}{\leq \overline{E}_{i}^\mathrm{CH}.\Gamma_{t,e,i} \quad {\forall t \in T, \forall e \in E, \forall i \in S}},{\label{eq: Limitation of charging power}}
		\addConstraint{0 \leq X^{-}_{t,e,i}}{\leq \overline{E}_{i}^\mathrm{CH}.{\Pi_{t,e,i}} \quad {\forall t \in T, \forall e \in E, \forall i \in S}},{\label{eq: Limitation of discharging power}}
		\addConstraint{\sum\limits_{i=1}^{S}(\Pi_{t,e,i}+\Gamma_{t,e,i})}{\leq 1 \quad {\forall t \in T, \forall e \in E}},{\label{eq: one mode of charging/discharging}}
		%
		%
		\addConstraint{\sum\nolimits_{e \in E} ({{\Gamma}_{t,e,i}+\Pi_{t,e,i})}}{\leq \overline{N}_{i}^{CH} \quad {\forall t \in T, \forall i \in S}},{\label{eq: Limitation of chargers for charging and discharging}}
		\addConstraint{\begin{aligned}[t]
				\inf_{\mathbb P\in \Xi^{EV}}\mathbb P &\bigg[\rho_{t,i}^{+}.X^{+}_{t,e,i}\leq \vartheta_{e}.\Gamma_{t,e,i}+\widehat{\rho}^{+}_{t,i}.\overline{E}_{e}.\Gamma_{t,e,i}\\&.\bigg(\widetilde{SOC}_{e}-e^{T}\omega_e+\widehat{A}_{t,e,i}-\frac{\mathcal{\widehat{O}}_{t,e,i}.\gamma_e}{\overline{E}_e}\bigg)\\& -\widehat{\rho}^{+}_{t,i}. \overline{E}_{e}.(1-\Gamma_{t,e,i}).\sum\limits_{t=1}^{t}{\frac{\zeta_{t,e}.\gamma_e}{\overline{E}_{e}}}\bigg]\geq1-\epsilon_{th}^{EV}\\&
				{\forall t \in T, \forall e \in E, \forall i \in S},
		\end{aligned}}{\label{eq: CC of Preference cost}}
		\addConstraint{\begin{aligned}[t]
				\inf_{\mathbb P\in \Xi^{EV}}\mathbb P & \bigg[\rho_{t,i}^{-}.X^{-}_{t,e,i}\geq\mathcal{G}_{e}.\Pi_{t,e,i}+\widehat{\rho}^{-}_{t,i}.\overline{E}_{e}.\Pi_{t,e,i}\\&.\bigg(\widetilde{SOC}_{e}-e^{T}\omega_e+\widehat{A}_{t,e,i}- \frac{\mathcal{\widehat{O}}_{t,e,i}.\gamma_e}{\overline{E}_{e}}\bigg)\\&-\widehat{\rho}^{-}_{t,i}.\overline{E}_{e}.(1-\Pi_{t,e,i}).\sum\limits_{t=1}^{t}{\frac{\zeta_{t,e}.\gamma_e}{\overline{E}_{e}}}\bigg]\geq1-\epsilon_{th}^{EV}\\&
				{\forall t \in T, \forall e \in E, \forall i \in S},
		\end{aligned}}{\label{eq: CC of Preference revenue}}
		\addConstraint{\Gamma_{t,e,i}.(\mathcal{O}_{t,e,i}+\mathcal{D}_{t,e,i})}{\begin{aligned}[t]\leq(\widehat{\mathcal{DO}}_{t,e}
				+\mathcal{K}_{e}).\Gamma_{t,e,i}\\~\mathllap
				{\forall t \in T, \forall e \in E, \forall i \in S},\end{aligned}}{\label{eq: Distance preference for charging}}
		\addConstraint{\Pi_{t,e,i}.(\mathcal{O}_{t,e,i}+\mathcal{D}_{t,e,i})}{\begin{aligned}[t]\leq(\widehat{\mathcal{DO}}_{t,e} 
				+\mathcal{K}_{e}).{\Pi_{t,e,i}}\\~\mathllap{\forall t \in T, \forall e \in E, \forall i \in S},\end{aligned}}{\label{eq: Distance preference for discharging}}
		\addConstraint{X^{+}_{t,e,i}}{\begin{aligned}[t]=0\quad\quad{\forall t \in T, \forall e \in E, \forall i=VCS},\end{aligned}}{\label{eq: charging power of VCS}}
		\addConstraint{X^{-}_{t,e,i}}{\begin{aligned}[t]=0\quad\quad{\forall t \in T, \forall e \in E, \forall i=VCS},\end{aligned}}{\label{eq: discharging power of VCS}}
		\addConstraint{\mathcal{O}_{t,e,i}+\mathcal{D}_{t,e,i}}{\begin{aligned}[t]=\zeta_{t,e}{\forall t \in (T-F_1-F_2), \forall e \in E, \forall i=VCS},\end{aligned}}{\label{eq: Distance to VCS for mandatory trip}}
		\addConstraint{\mathcal{O}_{t,e,i}+\mathcal{D}_{t,e,i}}{\begin{aligned}[t]=0\quad{\forall t \in F_1, \forall e \in E, \forall i=VCS},\end{aligned}}{\label{eq: Distance to VCS for the first optional trip}}
		\addConstraint{\mathcal{O}_{t,e,i}+\mathcal{D}_{t,e,i}}{\begin{aligned}[t]=0\quad{\forall t \in F_2, \forall e \in E, \forall i=VCS}.\end{aligned}}{\label{eq: Distance to VCS for the second optional trip}}
	\end{mini!}

	\subsection{Joint DRCC Model in CS Layer}
	\label{subsec:DRCC in CS}
	The DRCC model in the CS layer is presented in this section as a maximization of the net revenue of all CSs. The sum of the CSs' revenue is the objective function of the DRCC problem in the CS layer as in Eq.~\eqref{eq: Net revenue of CSs}. The revenue of CS $i$ is derived from selling electricity to EV $e$ and the aggregator during G2V and V2G operation, respectively. The costs of CS $i$ include the operational cost of CGU, cost of electricity purchased from retailer $r$ and EV $e$ in G2V and V2G services, respectively. 
	
	\begin{maxi!}|s|[3]<b>
		{\substack{Y^{re}_{t,i,r},Y^{GU}_{t,i},\widetilde{Y}^\mathrm{PV}_{t,i} \\ Y^+_{t,i},Y^-_{t,i},\rho_{t,i}^{-},\\\beta_{t,i,r},\psi_{t,i}
		}}{\begin{aligned}[t]\sup_{\mathbb P\in \Xi^{CS}}\mathbb E_\mathbb P\bigg[{\sum\limits_{t=1}^{T}\sum\limits_{i=1}^{S}}\sum\nolimits_{e \in E} (X^{+}_{t,e,i}.\rho_{t,i}^{+}\\~+X^{-}_{t,e,i}.\rho_{t,i}^\mathrm{AG})\\~\mathllap{-Y^\mathrm{re}_{t,i,r}.\rho_{t,r}^\mathrm{re}-\sum\nolimits_{e \in E}X^{-}_{t,e,i}.\rho_{t,i}^{-}}\\~\mathllap{-\frac{Y^\mathrm{GU}_{t,i}.\rho_{t}^\mathrm{gas}}{\eta_{i}^\mathrm{GU}.HV}}\bigg]{\quad \quad \quad \quad \quad \forall r \in R},\end{aligned}}{\label{eq: Net revenue of CSs}}{}
		\addConstraint {\begin{aligned}[t]\inf_{\mathbb P\in \Xi^{CS}}\mathbb P&\bigg[{\widetilde{Y}^\mathrm{PV}_{t,i}-\textbf{1}^\mathrm T\xi_{t,i}+Y^\mathrm{GU}_{t,i}+ Y^\mathrm{re}_{t,i,r}+ Y^-_{t,i}+\sum\nolimits_{e \in E} X^{-}_{t,e,i}}\\&+\varepsilon \geq  \frac{\sum\nolimits_{e \in E} X^{-}_{t,e,i}}{\eta_i^\mathrm{CH}}{+\frac{\sum\nolimits_{e \in E} X^{+}_{t,e,i}}{\eta_i^\mathrm{CH}}+ Y^{+}_{t,i}\bigg]}\geq 1-\epsilon_{th}^{CS}\\&{\forall t \in T, \forall i \in S, \forall r \in R},\end{aligned}}{\label{eq: CC of power balance in CSs}}
		\addConstraint{0\leq}{Y^\mathrm{GU}_{t,i} \leq \overline{E}^\mathrm{GU}_{i}\quad{\forall t \in T,\forall i \in S}},{\label{eq: CGU capacity constraint}}
		\addConstraint\inf_{\mathbb P\in \Xi^\mathrm{CS}}\mathbb P \bigg[{0\leq}{\widetilde{Y}^\mathrm{PV}_{t,i}-\textbf{1}^\mathrm T\xi_{t,i} \leq \overline{E}^\mathrm{PV}_{i}\bigg]\geq 1-\epsilon_{th}^{CS}\quad{\forall t \in T,\forall i \in S}},{\label{eq: CC of PV capacity constraint}}
		\addConstraint {\begin{aligned}[t]\inf_{\mathbb P\in \Xi^\mathrm{CS}}\mathbb P&\bigg[\underline{r}_i^\mathrm{PV} \leq \frac{\widetilde{Y}_{t,i}^\mathrm{PV}-\widetilde{Y}_{t-1,i}^\mathrm{PV}-\widehat{\xi}_{t,i}}{\Delta t}
				\leq \overline{r}_i^\mathrm{PV}\bigg]\geq 1-\epsilon_{th}^\mathrm{CS},\end{aligned}}{\label{eq: CC of spatial correlation}}
		\addConstraint{0\leq}{Y^\mathrm{re}_{t,i,r} \leq \overline{E}_{i}.\beta_{t,i,r}\quad{\forall t \in T, \forall i \in S, \forall r \in R}},{\label{eq: Limitation of power purchased from retailers}}
		\addConstraint{\sum\limits_{r=1}^{R} \beta_{t,i,r}}{\leq 1\quad{\forall t \in T, \forall i \in S}},{\label{eq: selection of retailers}}
		\addConstraint{0\leq Y^+_{t,i}}{\leq \overline{E}_{i}^\mathrm{ESS}.\psi_{t,i}\quad{\forall t \in T, \forall i \in S}},{\label{eq: Limitation of charging ESS}}
		\addConstraint{ 0\leq Y^{-}_{t,i}}{\leq \overline{E}_{i}^\mathrm{ESS}.(1-\psi_{t,i})\quad{\forall t \in T, \forall i \in S}},{\label{eq: Limitation of discharging ESS}}
		\addConstraint{\underline{SOC}_{i}^\mathrm{ESS} \leq}{\frac{ \sum\limits_{t=2}^{t}(Y^+_{t,i}-Y^-_{t,i}).\Delta t}{\overline{E}_{i}^{ESS}}\leq \overline{SOC}_{i}^\mathrm{ESS}\quad{\forall t \in T, \forall i \in S}},{\label{eq: Limitation of SOC of ESS}}
		\addConstraint{\underline{\rho}^{-}\leq}{\rho_{t,i}^- \leq \overline{\rho}^{-}\quad{\forall t \in T, \forall i \in S}}.{\label{eq: Limitation of discharging price}}
	\end{maxi!}

	There exists a set of DRCC as in Eqs.~\eqref{eq: CC of power balance in CSs},~\eqref{eq: CC of PV capacity constraint},~\eqref{eq: CC of spatial correlation} in this layer. The probability of maintaining power balance between supply and demand at CS $i$ under the worst distribution in ambiguity set for CS layer ($\Xi^{CS}$) is imposed by Eq.~\eqref{eq: CC of power balance in CSs}. The lower and upper capacity limit of CGU are enforced by Eq.~\eqref{eq: CGU capacity constraint}. The DRCC for ensuring the lower and upper capacity limit of PV generation at CS $i$ is given in Eq.~\eqref{eq: CC of PV capacity constraint}. Equation~\eqref{eq: CC of spatial correlation} represents temporal correlation between renewable energy generation uncertainty by restricting the ramping of PV generation. For this constraint, the uncertainty parameter contains both $\xi_{t,i}$ and $\xi_{t-1,i}$ as: \par
	\begin{equation}
		\widehat{\xi}_{t,i}=
		\begin{bmatrix}
			\xi_{t,i}\\
			\xi_{t-1,i}
		\end{bmatrix}.
	\end{equation}
	
	In order to consider the effect of temporal correlation on Eq.~\eqref{eq: CC of spatial correlation}, covariance matrix should be defined as \cite{pourahmadi2019distributionally}:
	\begin{equation}
		\widehat{\Sigma}_{t,i}^\mathrm{PV}=
		\begin{bmatrix}
			\Sigma_{t,i}^\mathrm{PV} & \Upsilon_{(t,t-1),i}^\mathrm{PV}\\
			\Upsilon_{(t,t-1),i}^\mathrm{PV} & \Sigma_{t-1,i}^{\mathrm {PV}}
		\end{bmatrix}.
	\end{equation}
	
	The electricity purchased from retailer $r$ is constrained by Eq.~\eqref{eq: Limitation of power purchased from retailers}. Equation~\eqref{eq: selection of retailers} guarantees that only one retailer is selected by CS $i$ at time $t$. The limitation of charging and discharging power of ESS at CS $i$ are established by Eqs.~\eqref{eq: Limitation of charging ESS} and \eqref{eq: Limitation of discharging ESS}. The SOC of ESS must be within a permissible range, which is enforced by Eq.~\eqref{eq: Limitation of SOC of ESS}. Equation~\eqref{eq: Limitation of discharging price} ensures that the electricity prices in V2G services is limited by minimum and maximum bounds for the DRCC problem in CS layer.

	\subsection{DRCC Model in Retailer Layer}
	The DRCC model in Retailer layer is presented in this section as a maximization of the net revenue of all retailers. It consists of the difference between revenue obtained by selling electricity to CS $i$, and the cost of electricity purchased from the wholesale electricity market, as given in Eq.~\eqref{eq: Revenue of retailers}. The wholesale electricity prices is the stochastic parameter in this layer.
	\label{subsec:DRCC in Retailer}
	\begin{maxi!}|s|[3]<b>
		{\substack{\widetilde{\rho}_{t,r}^\mathrm{re}}}{\begin{aligned}[t]\sup_{\mathbb P\in \Xi^{CS}}\mathbb E_\mathbb P\bigg[{\sum\limits_{t=1}^{T}\sum\limits_{r=1}^{R}}&\sum\nolimits_{i \in {S}}{Y^{re}_{t,i,r}}.{{\rho}^\mathrm{re}_{t,r}}\\&
				-P^\mathrm{WM}_{t,r}.(\widetilde{\rho}_{t}^\mathrm{WM}-\textbf{1}^T\kappa_{t})\bigg]\quad{\forall i \in S,
		}\end{aligned}}{\label{eq: Revenue of retailers}}{}
		\addConstraint{P_{t,r}^\mathrm{WM}}{=\sum\nolimits_{i \in S}{Y^{re}_{t,i,r}},
		}{\label{eq: active power balance of DN}}
		\addConstraint{Q_{t,r}^\mathrm{WM}}{=\sum\nolimits_{i \in S}  Q^{re}_{t,i,r},
		}{\label{eq: reactive power balance of DN}}
		\addConstraint{P_{m,n,t}}{\begin{aligned}[t]=g_{m,n}.(1+\Delta\hat{V}_{m,t}).(\Delta V_{m,t}-\Delta V_{n,t})\\~\mathllap{-b_{m,n}.(\theta_{m,t}-\theta_{n,t})\quad{\forall m,n \in B, \forall t \in T}},\end{aligned}}{\label{eq: active power line}}
		\addConstraint{Q_{m,n,t}}{\begin{aligned}[t]=-b_{m,n}.(1+\Delta\hat{V}_{m,t}).(\Delta V_{m,t}-\Delta V_{n,t})\\~\mathllap{-g_{m,n}.(\theta_{m,t}-\theta_{n,t})\quad{\forall m,n \in B, \forall t \in T }},\end{aligned}}{\label{eq: reactive power line}}
		\addConstraint{V_{m,t}}{=1+\Delta V_{m,t}\quad{\forall m \in B, \forall t\in T}},{\label{eq: voltage}}
		\addConstraint{\theta_{m,t}}{=0+\Delta \theta_{m,t}\quad{\forall m \in B,\forall t\in T}},{\label{eq: angle}}
		\addConstraint{\underline{\Delta V}_m\leq}{\Delta V_{m,t} \leq \overline{\Delta V}_m
			\quad{\forall m \in B}},{\label{eq: Limitation of voltage}}
		\addConstraint{\underline{P}_{m,n}\leq}{P_{m,n,t} \leq \overline{P}_{m,n}
			\quad{\forall m,n \in B, \forall t \in T}},{\label{eq: Active power limitation}}
		\addConstraint{\underline{Q}_{m,n}\leq}{Q_{m,n,t} \leq \overline{Q}_{m,n}
			\quad{\forall m,n \in B, \forall t \in T}},{\label{eq: Reactive power limitation}}
		%
		%
		\addConstraint {\begin{aligned}[t]\inf_{\mathbb P\in \Xi^{re}}\mathbb P&\bigg[\underline{\alpha}_t \times \widetilde{\rho}^\mathrm{WM}_{t}-\textbf{1}^T\kappa_t\leq {{\rho}^\mathrm{re}_{t,r}} \leq \overline{\alpha}_t \times \widetilde{\rho}^\mathrm{WM}_{t}-\textbf{1}^T\kappa_t
				\bigg] \\&\geq  1-\epsilon_{th}^{re} .\end{aligned}}{\label{eq: CC of limitation of retailer prices}}
	\end{maxi!}
	
	Equations \eqref{eq: active power balance of DN} and \eqref{eq: reactive power balance of DN} enforce the active and reactive power balance at all times, respectively. Thus, the electricity purchased from the wholesale electricity market through retailer $r$ must be equal to sum of the electricity purchased by CSs from retailer $r$ for active and reactive power at time $t$. Equations \eqref{eq: active power line} and \eqref{eq: reactive power line} satisfy real and reactive power flows in the network considering voltage magnitude and angle deviations determined by Eqs.~\eqref{eq: voltage} and \eqref{eq: angle}. Equation \eqref{eq: Limitation of voltage} enforces the node voltages between minimum and maximum bounds. Equations~\eqref{eq: Active power limitation} and \eqref{eq: Reactive power limitation} guarantee that active and reactive power are maintained within a standard range. The probability of maintaining the electricity prices offered by retailers between maximum and minimum bounds under the worst distribution ambiguity set in Retailer layer ($\Xi^{re}$) is imposed by Eq.~\eqref{eq: CC of limitation of retailer prices}.
	
	\section{Simulation Results}
	\label{sec: simulation results}
\begin{figure}[!htb]
		\setlength\abovecaptionskip{-0.5\baselineskip}
		\centering
		\includegraphics[ width=1\columnwidth]{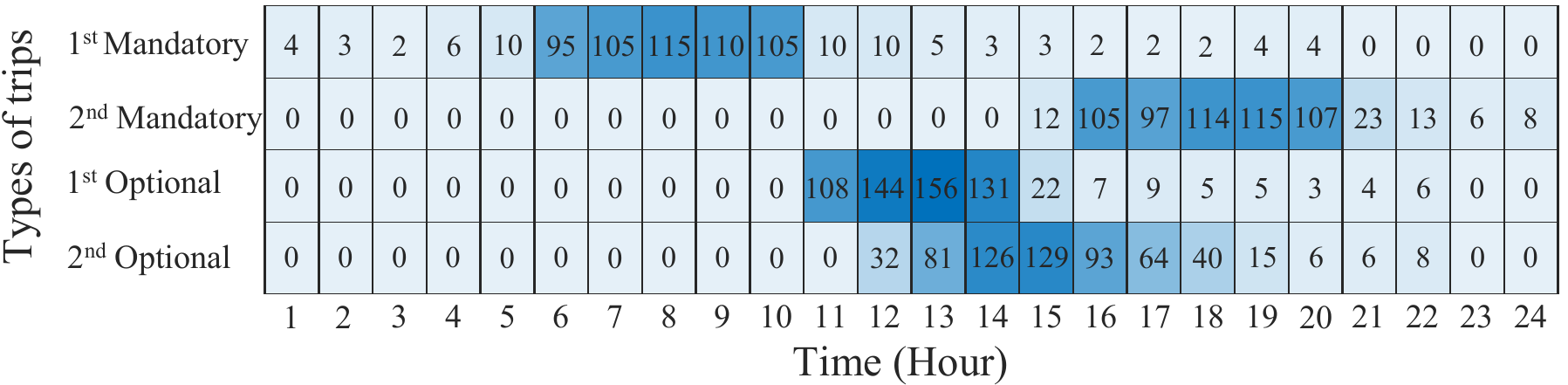}
		\caption{Number of EVs in different trips}
		\label{fig: number_of_EV_Heatmap}
		\captionsetup{justification=centering}
	\end{figure}

	We implemented the proposed day-ahead DRCC scheduling framework in a simulation study with three retailers, nine CSs, and 600 EVs in an small area of San Francisco, the USA, using IEEE 37-node distribution test system to assess its performance under different conditions. 30 bidirectional fast DC chargers with the capacity of 50 kW are considered in all CSs. Each CS is equipped with a 65 kW CGU, a PV system with a capacity of $\{16,19.2,24,27.2,32\}$ kW and one-hour ESS with the capacity of $\{45,50,65,70,85\}$ kW. PV systems and ESS sizes are randomly assigned to CSs. The mean value of the initial SOC of 600 EVs is between 10\% and 95\% and the covariance of initial SOC of each EV is equal to 5\%. It is assumed that each EV plans two mandatory trips and two optional trips in a typical day. 
	As depicted in Fig.~\ref{fig: number_of_EV_Heatmap}, the first mandatory trip of 88.3\% of EVs in the fleet is randomly scheduled between 06:00 to 10:00. The first optional trip of 93.5\% of EVs is randomly planned between 11:00 to 15:00. The second optional trip of 88.8\% of EVs is assumed to occur between 13:00 to 18:00. Finally, the second mandatory trip of 89.7\% of EVs is supposed to take place between 16:00 to 20:00. Before solving the scheduling problem, the shortest driving routes between the origin of EV $e$, the location of CS $i$, and the destination of EV $e$ for each trip are determined by ArcGIS\textsuperscript{\textregistered}.  
	The ambiguity sets for PV generation and wholesale electricity prices, which include a family of probability distribution functions with the same mean and covariance, are constructed from historical data. We calculate the mean value of PV generation ($\widetilde{Y}^{PV}_{t,i}$) as well as the covariance matrix of PV generation ($\xi_{t,i}$ and $\widehat{\xi}_{t,i}$) for each hour from data of 100 days extracted from Renewables.ninja for the same area in San Francisco \cite{ninja}. In addition, for calculating the mean value ($\widetilde{\rho}_t^\mathrm{WM}$) and covariance ($\kappa_t$) of the wholesale electricity prices, 100 days worth of prices are used from California ISO \cite{californiaISO}. To obtain the prices offered to CSs by the retailers, the day-ahead electricity prices of the wholesale market is multiplied by 4.5 homogeneously in order to consider network maintenance costs, ancillary services costs, taxes etc. The profit margin of the retailers ($\underline{\alpha}$ and $\overline{\alpha}$) is assumed to be 5-50\% in the G2V operation, while the CSs seek profit in the range of 10\% to 30\% of the true energy prices. Furthermore, during the V2G service, electricity prices offered by CSs are between 60-85\% less than the prices offered by the retailers. The electricity prices sold to the aggregator by CSs is 10\% more than what CSs offered to EV drivers for V2G service.
	
	\subsection {Evaluation of Low- to High-Risk Cases }
	First, we solved the DRCC problem in different layers for different confidence levels ($\nu=1-\epsilon$) to investigate the changes in cost and profits of the stakeholders. For the low-risk (conservative) case study, the confidence level is high and the constraints in Eqs.~\eqref{eq: CC of SOC of EVs}, \eqref{eq: CC of SOC at the end of a day}, \eqref{eq: CC of Preference cost}, and \eqref{eq: CC of Preference revenue} in EV layer, and Eqs.~\eqref{eq: CC of power balance in CSs}, ~\eqref{eq: CC of PV capacity constraint}, and ~\eqref{eq: CC of spatial correlation} in CS layer will be satisfied 95\% of the time or more. In the same case study, Eq.~\eqref{eq: CC of limitation of retailer prices} in retailer layer will be considered with a probability that is higher than or equal 90\% because the DRCC model will be infeasible at 95\% confidence level. It means that the retailers cannot supply CSs under the most conservative condition of the entire ecosystem. This would have not been revealed if we hadn't consider the impact of different layers on each other in the proposed iterative model. 
	
\begin{figure}[!htb]
		\setlength\abovecaptionskip{-0.5\baselineskip}
		\centering
		\makebox[\linewidth][c]
		{
			{\includegraphics[width=0.5\textwidth]{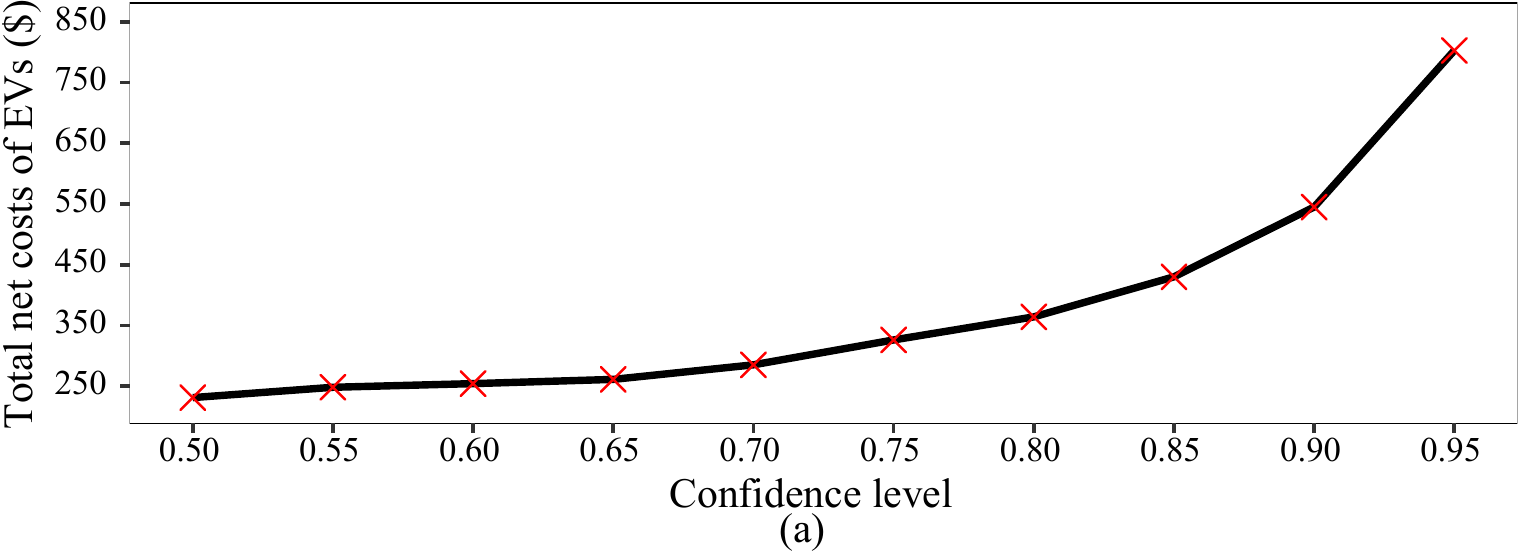}}}{   
			{\includegraphics[width=0.5\textwidth]{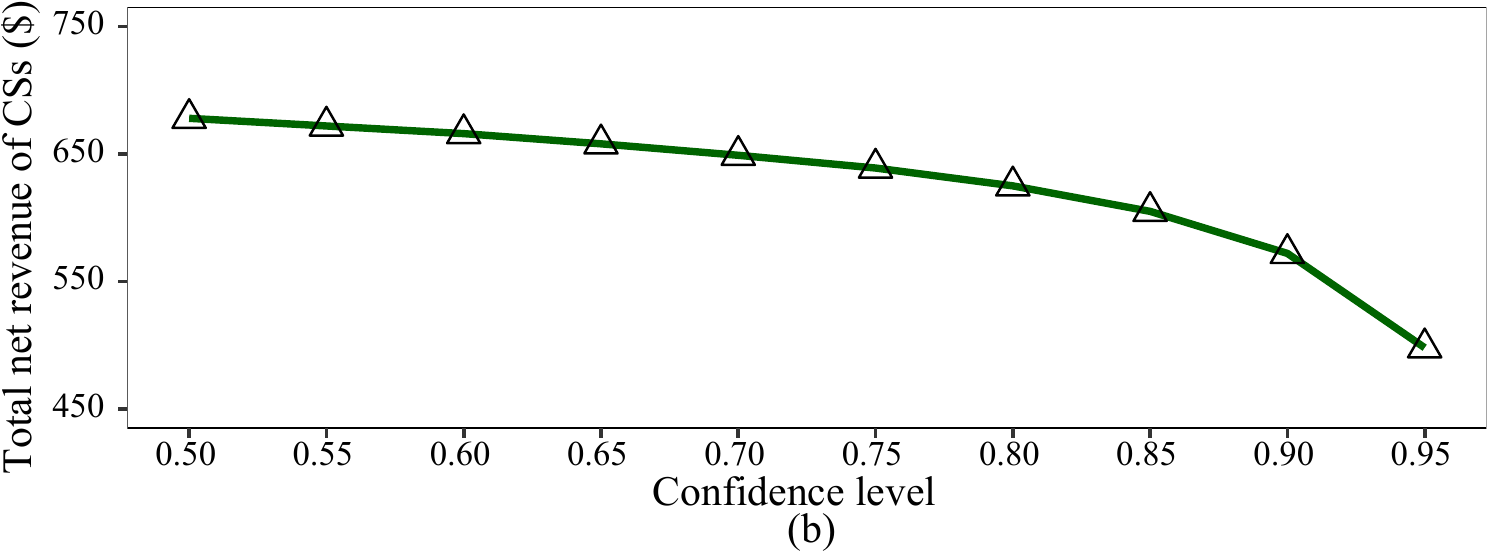}}  
		}{   
			{\includegraphics[width=0.5\textwidth]{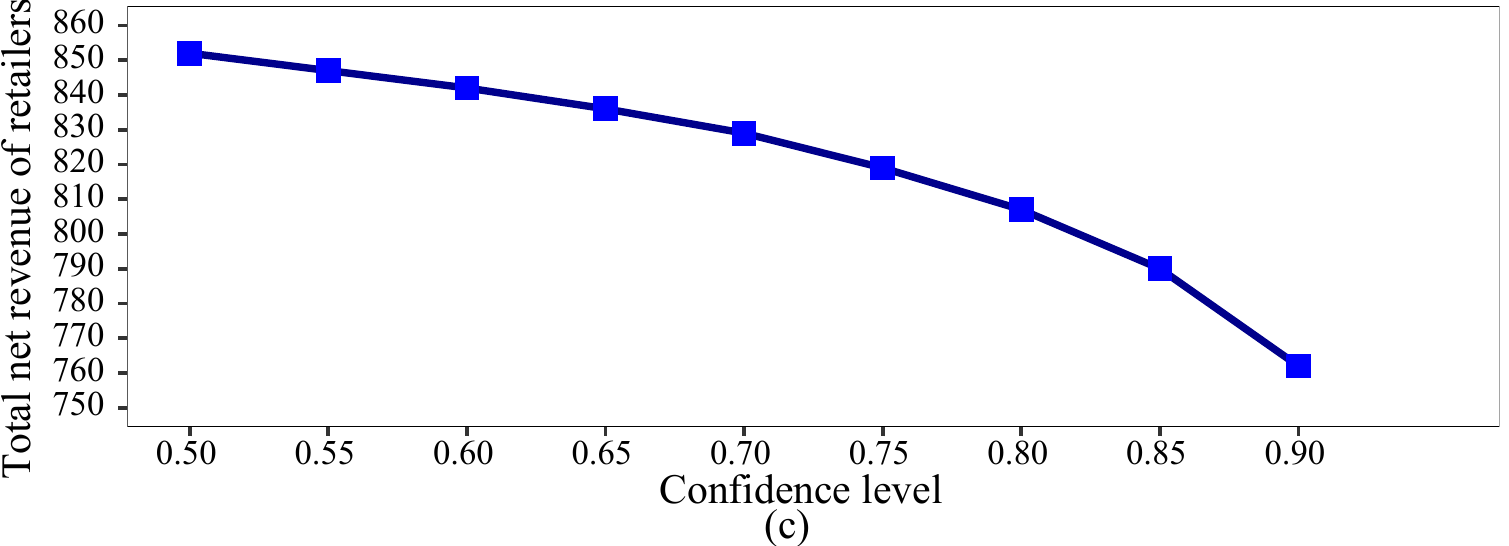}}  
		}
		\caption{(a) Total net cost of EVs, (b) total net revenue of CSs, and (c) total net revenue of retailers for different confidence levels from 0.5 (high-risk case) to 0.95 (low-risk case)}
		\label{fig: Convergence}
		\vspace{-0mm}
	\end{figure}

\begin{figure}[!htb]
		\setlength\abovecaptionskip{-0.5\baselineskip}
		\vspace*{-\baselineskip}
		\centering
		\includegraphics[width=1\columnwidth]{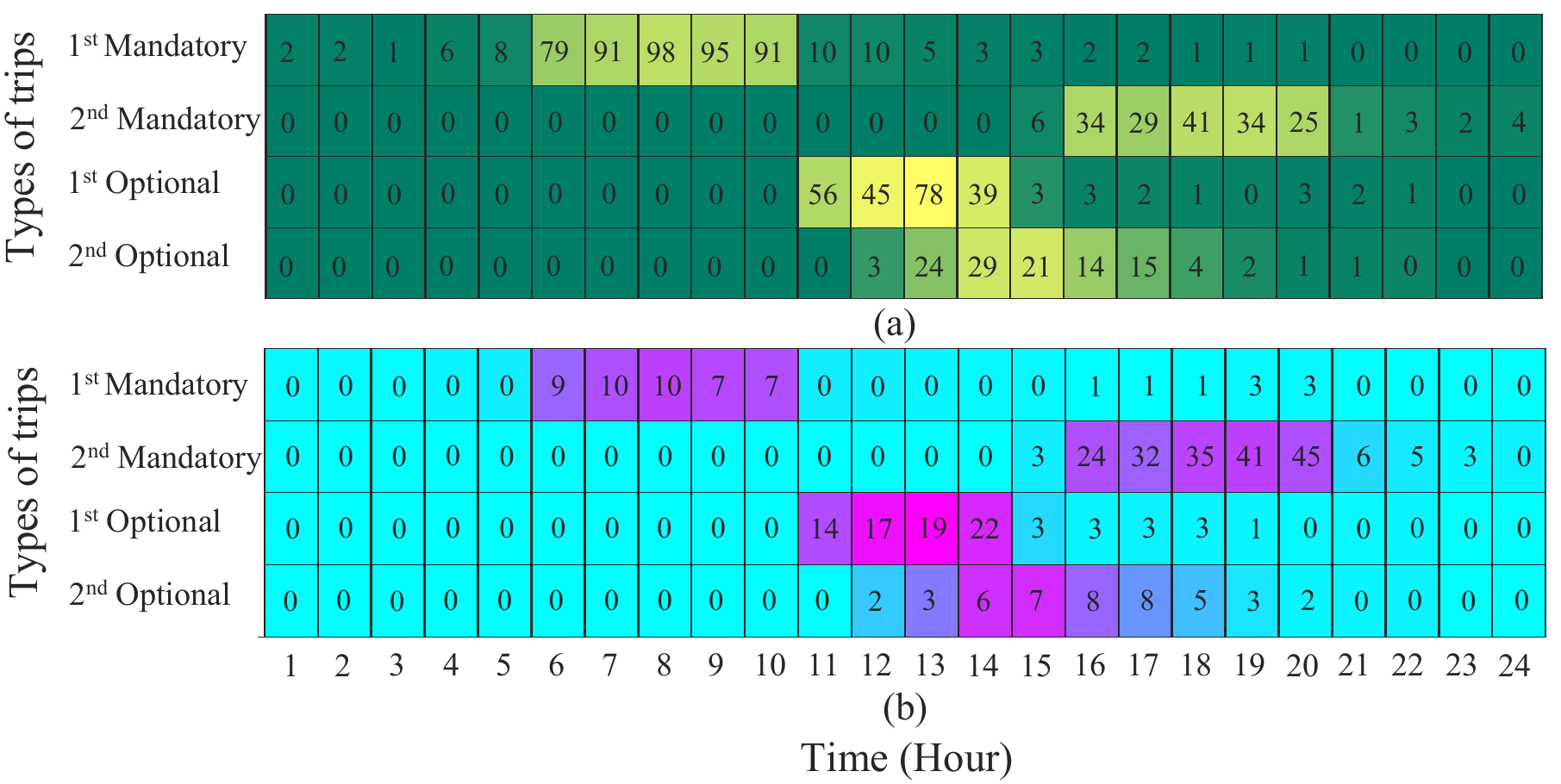}
		\caption{Number of EVs schedule for (a) G2V and (b) V2G operation}
		\label{fig: No_EV_G2V}
		\captionsetup{justification=centering}
	\end{figure}

	The total net cost of EVs, and the total net revenue of CSs and retailers are illustrated in Fig.~\ref{fig: Convergence} for different confidence levels. By increasing the confidence level from 0.5 (high-risk case study) to 0.95 (low-risk case study), the total net cost of EVs increased from \$231 to \$803 while the total net revenue of CSs and retailers decreased from \$678 and \$852 to \$498 and \$762, respectively. The reason is that at the lower confidence levels, constraints are relaxed for all stakeholders; hence more options are available to choose the most optimal V2G and G2V operation. Consequently, the risk of not meeting the day-ahead commitment for all stakeholders is much higher in this case in the real-time operation. However, at higher confidence levels, the feasible solution space is much smaller for all stakeholders. This way, they pay a premium for a higher confidence (lower risks) at the time of delivery of services. Since the DRCC problem for the three layers are solved iteratively, the impact of conservative operation of one layer leads other players to behave more conservatively, which resulted in disproportionately higher prices. 
	The number of EVs scheduled for G2V and V2G operations in each trip is shown in Fig.~\ref{fig: No_EV_G2V} for 0.95 confidence level in EV and CS layer, and 0.9 in retailer level. 
	
	\subsection {Validation of DRCC Formulation}
\begin{figure}[!htb]
		\setlength\abovecaptionskip{-0.5\baselineskip}
		
		\centering
		\includegraphics[ width=1\columnwidth]{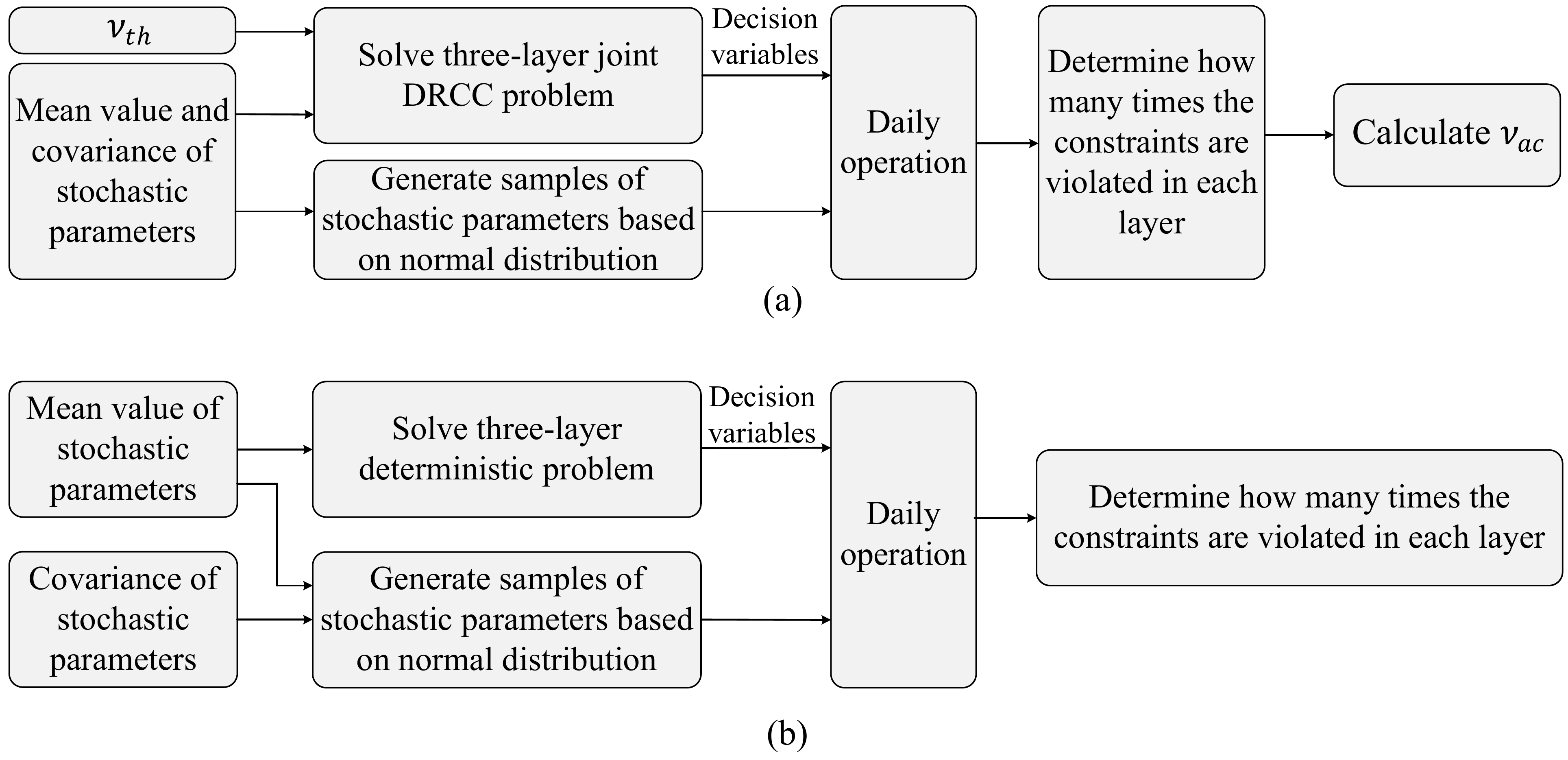}
		\caption{Flowchart of determining actual confidence level in (a) the DRCC problem and (b) the deterministic problem}
		\label{fig: flowchart_DRCC}
		\captionsetup{justification=centering}
	\end{figure}

	In this subsection, the quality of the proposed three-layer joint DRCC framework and the solutions are investigated. The DRCC solutions are valid while the actual confidence level ($\nu_{ac}=1-\epsilon_{ac}$) is more than or equal to the theoretical confidence level ($\nu_{th}=1-\epsilon_{th}$). For this investigation, as shown in Fig.~\ref{fig: flowchart_DRCC}(a), we firstly need to consider a theoretical confidence level, $\nu_{th}$, to solve the DRCC problems at different layers iteratively based on the mean value and covariance of stochastic parameters. 
	
	
	
	
	

	Afterwards, the optimal solutions jointly used with the samples created for the stochastic parameters 
	(representing the realised value of the parameters in real-time operation) to run a daily operation of the ecosystem. Then, we could check the number of times in which the CCs (Eqs.~\eqref{eq: CC of SOC of EVs}, ~\eqref{eq: CC of SOC at the end of a day}, ~\eqref{eq: CC of Preference cost}, and ~\eqref{eq: CC of Preference revenue} in EV layer, Eqs.~\eqref{eq: CC of power balance in CSs}, ~\eqref{eq: CC of PV capacity constraint}, and ~\eqref{eq: CC of spatial correlation} in CS layer, and Eq.~\eqref{eq: CC of limitation of retailer prices} in retailer layer) have been violated to obtain the actual confidence level, $\nu_{ac}$. 
	This process is repeated for different values of theoretical confidence level in each layer, i.e., $\nu_{th} \in [0.5, 0.9]$, to obtain the mean values of the actual confidence level, $\nu_{ac}$. While we used normal distribution in this part of our simulation study, it should be noted that the reformulation of the single-sided and double-sided CCs, presented in \cite{electronic}, is independent of the type of probability distribution function. Also, we calculated the CCs violations in a deterministic day-ahead scheduling framework using the process shown in Fig.~\ref{fig: flowchart_DRCC}(b). 
	
\begin{figure}[!htb]
		\setlength\abovecaptionskip{-0.2\baselineskip}
		\centering
		\includegraphics[width=1\columnwidth]{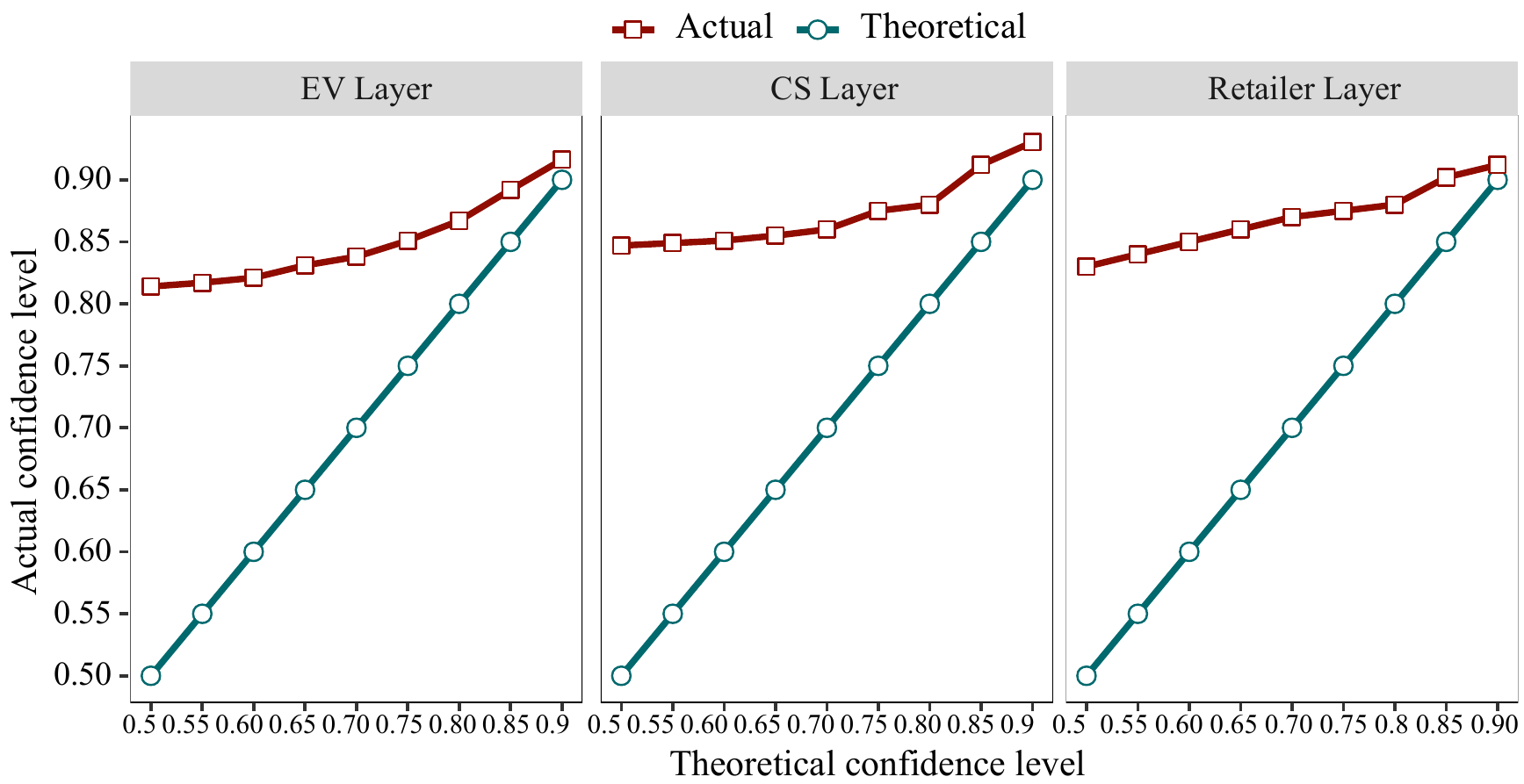}
		\caption{DRCC validation for EV, CS, and Retailer layers}
		\label{fig: CF}
		\captionsetup{justification=centering}
	\end{figure}

	The actual confidence levels obtained by the proposed stochastic framework are illustrated in Fig.~\ref{fig: CF} for each layer in comparison with the theoretical ones. Please note that for determining the actual confidence level of each layer, the theoretical confidence level of other layers are kept constant at 0.9. It can be seen from the figure that the actual confidence level is always higher than the theoretical one. In addition, Fig.~\ref{fig: CF} shows that the DRCC programming is more conservative on the lower range of theoretical confidence levels. 
	The simulation results for the deterministic scheduling framework show actual confidence levels of 0.71, 0.79, and 0.75 for EV, CS, and retailer layer, respectively, which are lower than the lowest actual confidence level of the proposed stochastic framework.

\begin{figure}[!htb]
		\setlength\abovecaptionskip{-0.2\baselineskip}
		\vspace*{-\baselineskip}
		\centering
		\includegraphics[width=1\columnwidth]{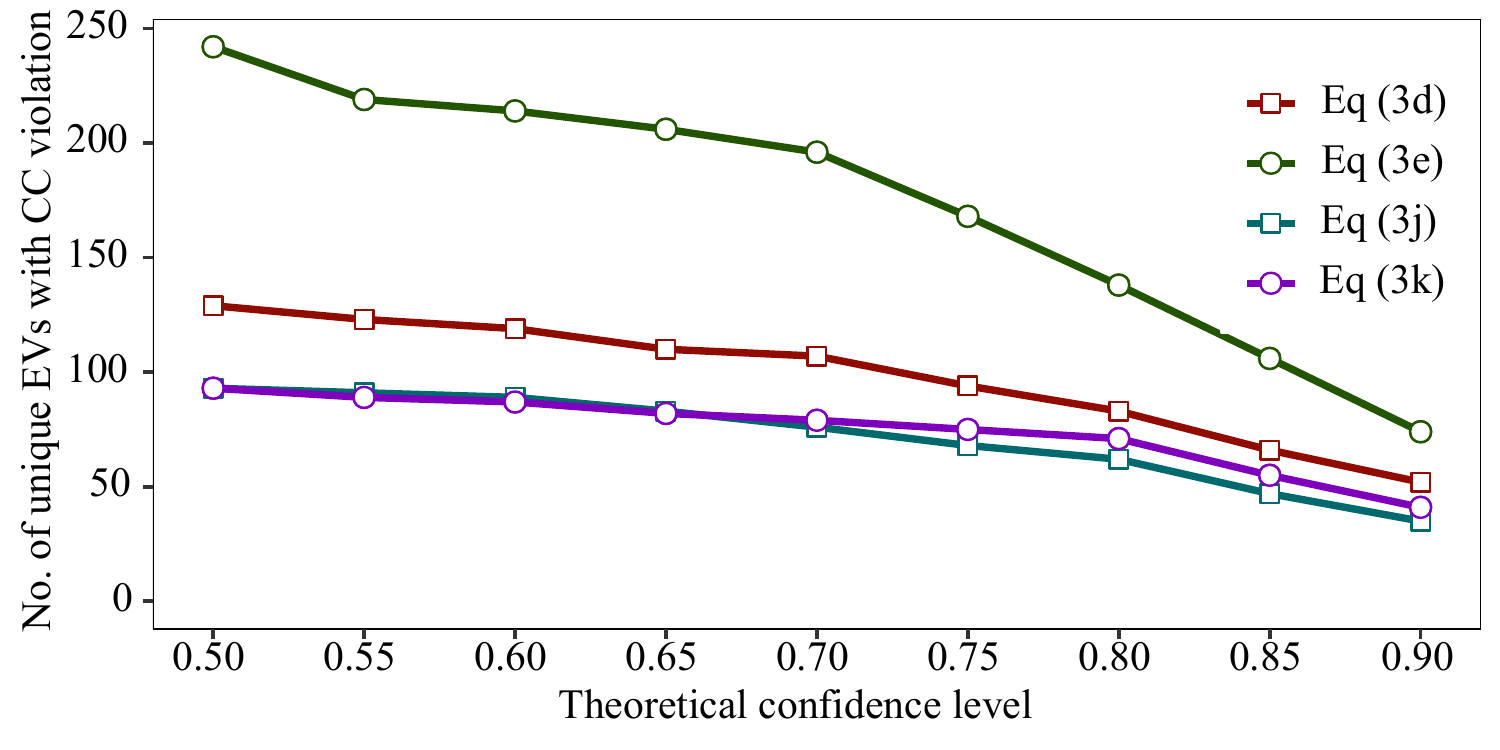}
		\caption{Number of unique EVs violating their CCs at least once a day at different confidence level in EV layer in the proposed DRCC framework.}
		\label{fig: No_EV}
		\captionsetup{justification=centering}
	\end{figure}

\begin{figure}[!htb]
		\setlength\abovecaptionskip{-0.2\baselineskip}
		\vspace*{-\baselineskip}
		\centering
		\includegraphics[width=1\columnwidth]{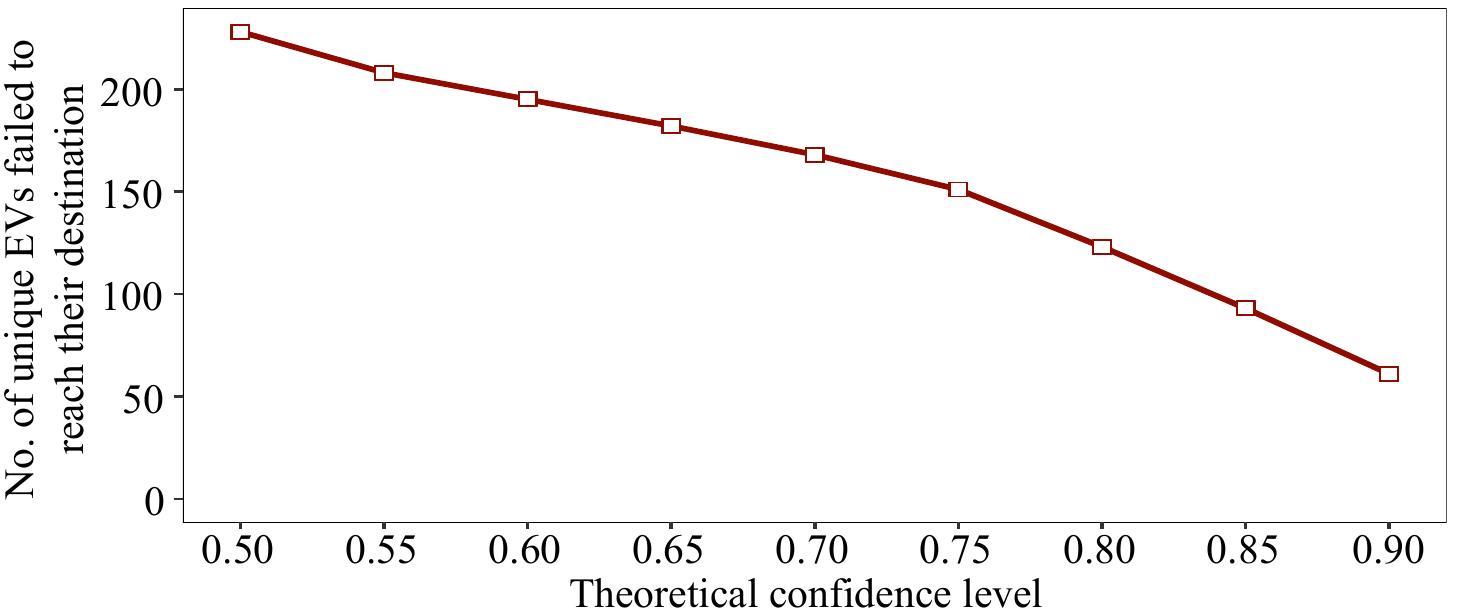}
		\caption{Number of unique EVs which does not fulfill their mandatory trips in the proposed DRCC framework at different confidence level.}
		\label{fig: No_EV_destination}
		\captionsetup{justification=centering}
	\end{figure}

	In addition, the number of unique EVs that violated the CCs (Eqs.~\eqref{eq: CC of SOC of EVs}, ~\eqref{eq: CC of SOC at the end of a day}, ~\eqref{eq: CC of Preference cost}, and ~\eqref{eq: CC of Preference revenue}) in the proposed DRCC framework are shown for different confidence levels in Fig.~\ref{fig: No_EV}. As expected, the number of unique EVs with constraint violation decreased by increasing the confidence level in EV layer. 
	Furthermore, Fig.~\ref{fig: No_EV_destination} depicts the number of unique EVs that does not reach their destination (due to lower SOC limit violation) at different confidence levels, which expectantly decreases by increasing the confidence level. It shows the importance of the proposed framework in reducing the EV drivers frustration, which contributes to lowering range anxiety. In addition, 272 unique EVs could not reach their destination in the deterministic problem, which is more than the one obtained at the lowest confidence level, 0.5, in the proposed DRCC framework.

	\subsection{Impact of Temporal Correlation of PV Generation}
\begin{figure}[!htb]
		\setlength\abovecaptionskip{-0.2\baselineskip}
		\vspace*{-\baselineskip}
		\centering
		\includegraphics[width=1\columnwidth]{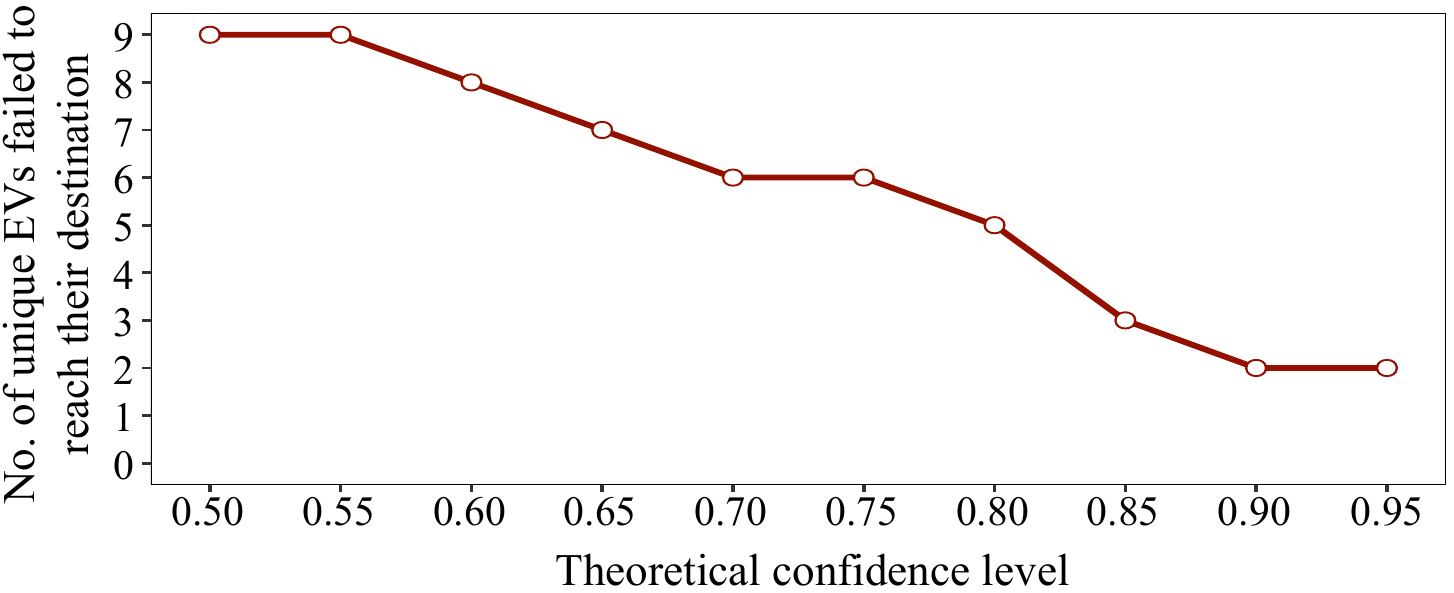}
		\caption{Number of unique EVs that could not reach their destination in the absence of PV correlation constraint at the CS layer.}
		\label{fig: No_EV_PV_Correlation}
		\captionsetup{justification=centering}
	\end{figure}

	In this section, the impact of considering the temporal correlation of PV generation in CSs are investigated. First, we calculated the root mean square error (RMSE) of PV generation with and without considering the temporal correlation in Eq.~\eqref{eq: CC of spatial correlation} for all CSs. It shows that the RMSE has improved from 17.8\% to 16.3\% by considering PV temporal correlations. 
	In addition, we calculated the number of unique EVs that could not fulfill their mandatory trip due to violating lower SOC limit at different confidence levels after removing the PV correlation effect in CS layer, which are shown in Fig.~\ref{fig: No_EV_PV_Correlation}. 
	It can be seen that an additional 2 EVs won't reach their destination at 0.95 confidence level. By decreasing the confidence level in the CS layer from 0.95 to 0.5 in the absence of PV correlation, the number of EVs that cannot fulfill their mandatory trips increases to 9. It shows the impact of PV temporal correlation on the successful scheduling of the ecosystem. 
	
	\section{Conclusion}
	\label{sec: conclusion}
	This paper proposes a three-layer joint DRCC model for the future e-mobility ecosystem including EVs, CSs, and retailers to schedule V2G and G2V services in the day ahead in an uncertain environment with unknown probability distribution functions. The interactions between stochastic parameters of the three stakeholders are considered in the proposed iterative model to improve the performance of the scheduling system for the entire e-mobility ecosystem. Also, second-order cone programming reformulation of the DRCC model is implemented to reformulate the double-sided CCs. In addition, the impact of temporal correlation of uncertain PV generation on the CSs operation is considered. The simulation results show that the choice of confidence level significantly affects the cost and revenue of the stakeholders as well as the accuracy of the schedules in real-time operation. For a low-risk case study, the model estimates 247.3\% increase in the total net cost of EVs compared to a high-risk case study, and a 26.6\% and 10.6\% decrease in the total net revenue of CSs and retailers, respectively. In addition, the number of unique EVs failed to reach their destination has decreased from 272 in deterministic scheduling model to 61 in low-risk case study. The simulation results prove the necessity of such planning frameworks to reduce the risks for all stakeholders, which in turn facilitates higher adoption of EVs by the end-users and investors. In future studies, we intend to explore different reformulation of DRCC in each layer that is less conservative at lower confidence levels. Also, to make the scheduling problem more flexible for the EV drivers, a new formulation will be developed to automatically select the best time for optional trips within a pre-defined range of time by the EV drivers.

	
	\renewcommand{\thesubsection}{\Alph{subsection}}
	\setcounter{section}{0}

	\bibliographystyle{IEEEtran}
	
	\bibliography{IEEEabrv,references}
	
\end{document}